\documentclass[pre,aps,twocolumn, floatfix, superscriptaddress,showpacs]{revtex4-1}
\usepackage{amsmath,bm,graphicx}
\usepackage{amssymb}
\usepackage{amsfonts}

\usepackage{graphicx}
\usepackage{hyperref}
\usepackage{latexsym}
\usepackage{color}
\usepackage[normalem]{ulem}

\def\be{\begin{equation}}
\def\ee{\end{equation}}
\def\bea{\begin{eqnarray}}
\def\eea{\end{eqnarray}}

\begin{document}
\title{Diffusion dynamics and steady states of  systems of hard rods on the square lattice}
\author{Saugata Patra}
\affiliation{Department of Physics, Indian Institute of Technology, Bombay, Powai, Mumbai-400076, India}
\author{Dibyendu Das}
\email{dibyendu@phy.iitb.ac.in}
\affiliation{Department of Physics, Indian Institute of Technology, Bombay, Powai, Mumbai-400076, India}
\author{R. Rajesh}
\email{rrajesh@imsc.res.in}
\affiliation{The Institute of Mathematical Sciences, CIT Campus, Taramani, 
Chennai-600113, India}
\affiliation{Homi Bhabha National Institute, Training School Complex, Anushakti Nagar, Mumbai 400094, India}
\author{Mithun K. Mitra}
\email{mithun@phy.iitb.ac.in}
\affiliation{Department of Physics, Indian Institute of Technology, Bombay, Powai, Mumbai-400076, India}

\begin{abstract}
 It is known from grand canonical simulations of a system of hard rods on two dimensional lattices that 
an orientationally ordered nematic phase exists only when the length of the rods is at least seven. However, a recent microcanonical simulation
with diffusion kinetics, conserving both total density and zero nematic order,  reported the existence of a nematically phase segregated steady 
state with interfaces in the diagonal direction for rods of length six [Phys. Rev. E {\bf 95}, 052130 (2017)], violating the  equivalence of different
ensembles for systems in equilibrium. We resolve this inconsistency by demonstrating that the kinetics violate detailed balance condition
and drives the system to a nonequilibrium steady state. By implementing diffusion kinetics that drives the system to equilibrium, 
even within this constrained ensemble, we recover earlier results showing phase segregation only for rods of length greater than or
equal to seven. Furthermore, in contrast to the nonequilibrium steady state, the interface has no  preferred orientational direction. In addition, by implementing different nonequilibrium kinetics, we show that the interface between the phase segregated states can lie in different directions depending on the choice of kinetics.
\end{abstract}

\pacs{64.70.mf,64.60.De,64.60.Cn,05.70.Fh}

\date{\today}
\maketitle

\section{Introduction}

In recent times, there has been considerable interest in the Statistical Mechanics of systems of long, hard rods on
lattices, also known as the Zwanzig model~\cite{1963-z-jcp-first}. The motivations include such systems being minimal 
models for studying entropy driven transitions like the isotropic-nematic transition~\cite{1949-o-nyas-effects,1956-f-rspa-statistical,1992-vl-rpp-phase} 
seen in more complicated
systems such as liquid crystals~\cite{1995-dp-oup-physics} and aqueous solutions of tobacco mosaic 
viruses~\cite{1989-fmcm-prl-isotropic}, as well as simple models   for studying adsorption of molecules
onto two dimensional substrates, experimental examples of which include 
oxygen monolayers adsorbed on molybdenum surfaces~\cite{1991-dmr-jcp-model}  and chlorine atoms adsorbed   
on silver~\cite{2001-mbr-ss-static}. More recently, interest in such models has been spurred by the demonstration of multiple entropy 
driven phase transitions in such simple lattices 
models of rods~\cite{2007-gd-epl-on} and rectangles~\cite{,2014-kr-pre-phase}.

Consider a  system  of non-overlapping $k$-mers, rods that occupy $k$ consecutive lattice sites along any one of the lattice directions. 
Though, early work based on virial expansion~\cite{1963-z-jcp-first}, high density expansions~\cite{1956-f-rspa-statistical} and the Guggenheim approximation~\cite{1961-d-jcp-statistics}, predicted a transition from the low-density disordered phase to a nematic-ordered phase with increasing density,
it can be shown that at densities close to full packing,  the nematic phase does not survive, as there are exponentially many disordered 
configurations~\cite{1995-dp-oup-physics,2007-gd-epl-on}.
It is only more recently the nature of the phases and phase transitions at high densities, as well as the minimum value of $k$
required for a nematic phase to exist, have become clearer through extensive Monte Carlo simulations. 
We first summarize the  results that are known for the model in two and three dimensions. Some rigorous results for the existence
of the nematic phase may be proved for certain values of $k$. In two dimensions, for $k=2$ (dimers), it is known  that the system is 
disordered at all densities~\cite{1972-hl-cmp-theory}, and that power law correlations exist only at  full 
packing~\cite{1963-fs-pr-statistical,2003-hkms-prl-coulomb}. The
existence of  a nematic phase, with orientational order and no positional order, at intermediate densities may be rigorously 
proved for very large $k$~\cite{2013-dg-cmp-nematic}. For intermediate values of $k$, Monte Carlo simulations show that for $k \geq 7$, 
the system undergoes two phase 
transitions: first from a low-density disordered phase to an intermediate density nematic phase, and second from the nematic
phase to a high density disordered phase~\cite{2007-gd-epl-on,2013-krds-pre-nematic,2013-kr-pre-reentrant}. The first transition belongs
to the Ising universality class on the square lattice~\cite{2008-mlr-epl-determination, 2008-mlr-jcp-critical,2009-fv-epl-restricted} and the
three state Potts model universality class on the triangular lattices~\cite{2008-mlr-jcp-critical,2008-mlr-pa-critical}. The nature of the
second transition has been difficult to resolve either numerically or analytically~\cite{2013-krds-pre-nematic,2013-kr-pre-reentrant}.  
On simpler tree-like lattices, the model may also be solved exactly. However, on such lattices, only one transition is
to a nematic phase is present for $k\geq 4$~\cite{2011-drs-pre-hard}, though models with soft interactions show two transitions~\cite{2013-kr-pre-reentrant}.
Density functional theory for rods give a similar result~\cite{2016-okdesh-jcp-monolayers}.
In three dimensions, the system has a richer phase diagram~\cite{2017-gkao-pre-isotropic,2017-vdr-xxx-different}. For $k>4$, the system shows a  
smectic-like order, where the densities of rods of one randomly chosen orientation is suppressed. For $k\geq 7$, a nematic phase is also obtained but for intermediate densities.

One particular question of interest is the smallest value of $k$, denoted by $k_{\min}$, required to observe a nematic phase. For rods in two dimensions,
from grand canonical Monte Carlo simulations, it is known $k_{\min}=7$~\cite{2007-gd-epl-on}. 
For rectangles of size $2 \times 2 k$ and $3 \times 3 k$, where $k$ is now
the aspect ratio, $k_{\min}$ is still $7$~\cite{2014-kr-pre-phase,2015-kr-epjb-phase,2015-kr-pre-asymptotic}. In three dimensions, large scale simulations of rods on cubic lattices show that $k_{\min}=7$~\cite{2017-gkao-pre-isotropic,2017-vdr-xxx-different}. This is in contrast
to $k_{\min}=4$ obtained from calculations based on  density functional theory~\cite{2016-okdesh-jcp-monolayers}, 
the Guggenheim approximation~\cite{1961-d-jcp-statistics}, as well as exact solutions on tree-like lattices~\cite{2011-drs-pre-hard}.
We note that the determination of $k_{\min}$ for two and three dimensional lattices are based on grand canonical Monte Carlo simulations where
both the total density as well as the nematic order parameter are allowed to fluctuate in time.

In a recent paper~\cite{2017-ltgv-pre-diffusiondriven}, the system of hard rods on the square lattice 
was studied in the micro canonical ensemble where both the total density as well as
the nematic order parameter were kept constant. The initial condition was prepared through random sequential adsorption of randomly oriented rods
till the ``jamming density" was achieved when no further rods could be adsorbed. The jamming density is dependent on $k$, while the nematic order
parameter is approximately zero for all $k$. Starting from this initial condition, the system is evolved by diffusion of rods in any of the four directions provided
the hard core constraint is not violated. Two interesting results were observed. First, for $k \geq 6$, the system phase separates into a horizontal
$k$-mer rich 
domain and a vertical $k$-mer rich domain at  large time scales. Second, the interface between them is at an angle $\pi/4$ to the horizontal, 
and this pattern of
diagonal stripes was shown to be an attractor for many different initial conditions. These results are surprising because in grand canonical 
simulations, nematic phase does not exist for $k=6$. However, phase separation in the micro canonical ensemble implies ordering in the grand canonical 
ensemble. Also, there is no a priori reason why the interface between the ordered phases for $k\geq 7$ should be along the diagonal. Simulations
in the grand canonical ensemble find interfaces in other directions including vertical and horizontal (for example see Fig.~2 of Ref.~\cite{2009-fv-epl-restricted}).
It would thus appear that diffusion seems to be driving the system to a state that is inconsistent with known results for the equilibrium phase in the grand canonical ensemble.

In this paper, we resolve these contradictions
by analysing more carefully the diffusion-driven self assembly of rods. In particular, we find that the nematic order for $k=6$ as well as the diagonal stripe pattern arises as a consequence of nonequilibrium nature of the diffusion kinetics as implemented in Ref.~\cite{2017-ltgv-pre-diffusiondriven}. 
By introducing a different diffusion dynamics that obey the detailed
balance condition, we show that there is no phase ordering for system with $k=6$ in the micro canonical ensemble. 
We also show that by introducing other nonequilibrium
dynamics, it is possible to obtain phase ordering  for $k=6$, but with varying patterns.

The remainder of the paper is organised as follows. In Sec.~\ref{methods}, we define the three models that we study in the paper. 
In Sec.~\ref{sec:results}, using 
Kolmogorov criteria, we show that the diffusion kinetics in Ref.~\cite{2017-ltgv-pre-diffusiondriven} does not obey detailed balance and hence
drives the system into a nonequilibrium steady state. We also present results for the steady state as well as 
approach to steady state for the new models introduced in this paper. Section~\ref{sec:conclusion} contains a summary and discussion
of results.

\section{Model \label{methods}}

Consider a system of hard $k$-mers, occupying $k$ consecutive lattice sites either in the horizontal or vertical direction, 
on a $L \times L$ square lattice. A lattice site may be occupied by at most one $k$-mer. We describe the dynamical model studied in Ref.~\cite{2017-ltgv-pre-diffusiondriven},  as well as  its generalisations that are
studied in this paper. The initial configuration is obtained through random sequential adsorption of $k$-mers, till it is not possible to deposit any
more $k$-mers. For each deposition, the orientation of the $k$-mer is chosen to be horizontal or vertical randomly. This procedure
results in configurations where number of vertical ($N_v$) and horizontal $k$-mers ($N_h$) are nearly equal, such that the 
nematic order parameter, $s = (N_v - N_h)/(N_v + N_h)$, has the value $s \approx 0$. Throughout this paper, we will consider density conserving
dynamical evolution of the system within this zero nematic order sector.  We note that this micro canonical ensemble is distinct from the
grand canonical simulations where both density and the nematic order parameter is allowed to fluctuate.

In the micro canonical ensemble with fixed nematic order parameter $s$, only translational diffusion moves 
are permitted, such that $N_h$ and $N_v$ are preserved throughout the simulation. In Ref.~\cite{2017-ltgv-pre-diffusiondriven}, the diffusion kinetics was implemented as follows. One of the $N_k = N_h + N_v$ $k$-mers is chosen at random. The $k$-mer may be shifted by one lattice spacing in any of the four directions, only if the move does not violate the hardcore constraints.  Let the number of allowed moves for the $k$-mer be denoted by $m$, where 
$m \leq 4$. One of these $m$ possible directions is chosen at random (with probability $1/m$) and the $k$-mer is shifted along that direction~\cite{2017-ltgv-pre-diffusiondriven}. The kinetics described above, where a $k$-mer shifts {\it randomly} (R) to any one of the available empty neighbouring sites shall be referred  to as the random nonequilibrium diffusion kinetics (NEDK-R). The nonequilibrium nature of the dynamics will be demonstrated in
Sec.~\ref{sec:detailed}.

Contrary to this dynamics, we define a random equilibrium diffusion kinetics (EDK) that satisfies detailed balance. In this kinetics, one of the 
$N_k$ $k$-mers is chosen randomly. This $k$-mer is then attempted to be shifted to one of its four possible neighbouring directions with equal probability. 
The attempted move succeeds provided the hard core constraint is not violated. If the move fails,  another $k$-mer is chosen at random.  For this kinetics, 
it is clear that all  moves are reversible and the  forward and backward transition probabilities of successful moves are equal to $(1/N_k).(1/4)$. 
Thus,  the EDK algorithm satisfies detailed balance and will lead to  a equilibrium steady state unless the system gets trapped in a metastable
nonequilibrium steady state.

We note that detailed balance can be violated in other ways than is done in the NEDK-R algorithm. For example, in this paper, we study the steady state for another nonequilibrium kinetics to demonstrate that the system may reach qualitatively different steady states from the ones found in 
Ref.~\cite{2017-ltgv-pre-diffusiondriven}. In this dynamics, a randomly chosen $k$-mer, if vertical, is attempted to be shifted in a clockwise sequence (North, East, South, West), while if horizontal, is attempted to be shifted in an anti-clockwise sequence (North, West, South, East). The first successful direction (respecting the hardcore constraint) is accepted to shift the $k$-mer. If the attempted shift in all the four directions fail, then we randomly choose 
another $k$-mer for an update. This combination of {\it clockwise-anticlockwise} (CAc) moves depending on the $k$-mer orientation will be
referred to as the NEDK-CAc algorithm.

The detailed results for the three dynamics is presented and contrasted with each other in Sec.~\ref{sec:results}. All the numerical results
presented in this paper are for a  $L \times L$ ($L = 256$) system. The density of occupied sites is given by $\rho = (k N_k)/L^2$, where
$N_k$ is the number of $k$-mers.  All the simulations are performed at the jamming density $\rho = \rho_j (k)$~\cite{2012-tll-pre-percolation}. 
We start with differently patterned initial conditions and determine if they lead to statistically similar steady state patterns after long times ($t \sim 10^8 - 10^9$), 
where time is measured in terms of a  Monte Carlo step comprising of  $N_k$ updates of rods.

\section{Results \label{sec:results}}

\subsection{NEDK-R dynamics violates detailed balance \label{sec:detailed}}

\begin{figure}
\centering
\includegraphics[width=\columnwidth]{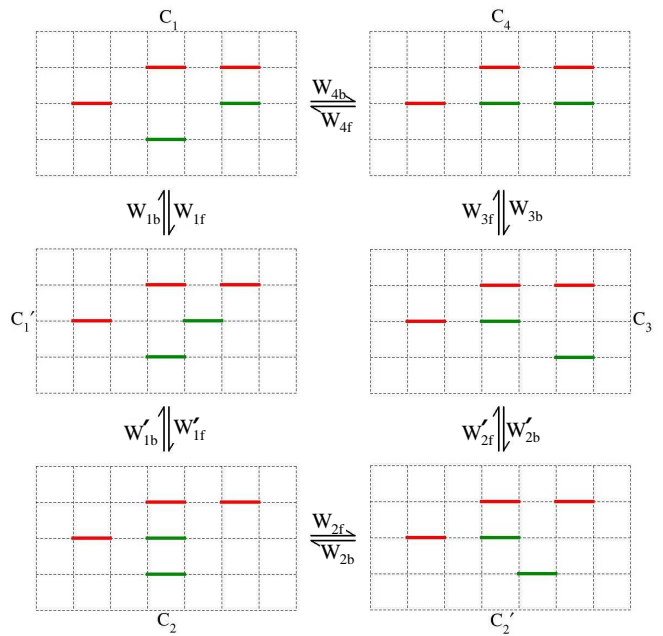}
\caption{\label{Fig1} Schematic diagram of $6$ configurations for illustrating the violation of detailed balance by the dynamics of
the model NEDK-R.}
\end{figure}

We first demonstrate that the NEDK-R dynamics violates detailed balance  and thus drives the system to a nonequilibrium
steady state. 
Consider the sequence of transitions  shown in Fig.~\ref{Fig1} denoting the cyclic transformation $C_1 \rightleftharpoons C_1' \rightleftharpoons C_2 \rightleftharpoons C_2' \rightleftharpoons 
C_3 \rightleftharpoons C_4 \rightleftharpoons C_1$. We will show that the rates for the forward and reverse
cycles are not  equal. For the transition $C_1 \rightleftharpoons C_1'$, the forward transition probability is $W_{1f} = (1/N_k).(1/3)$ and the backward transition probability is $W_{1b} = (1/N_k).(1/2)$. The factor $1/N_k$ accounts for the probability of choosing the $k$-mer.
The second factor is the probability $1/m$ ($m$ being the number of allowed diffusion moves for the $k$-mer), 
with $m$ being distinct for the randomly chosen $k$-mer which is attempted to be moved. 
Similarly, the forward and backward probabilities for the full sequence of transitions shown in Fig.~\ref{Fig1} are
\begin{eqnarray*}
W_{1f} = (1/N_k).(1/3) ~~~&~~~ W_{1b} = (1/N_k).(1/2) \\
W'_{1f} = (1/N_k).(1/2) ~~~&~~~ W'_{1b} = (1/N_k).(1/1) \\
W_{2f} = (1/N_k).(1/3) ~~~&~~~ W_{2b} = (1/N_k).(1/3) \\
W'_{2f} = (1/N_k).(1/3) ~~~&~~~ W'_{2b} = (1/N_k).(1/4) \\
W_{3f} = (1/N_k).(1/4) ~~~&~~~ W_{3b} = (1/N_k).(1/2) \\
W_{4f} = (1/N_k).(1/1) ~~~&~~~ W_{4b} = (1/N_k).(1/4). \\
\end{eqnarray*}
Thus,  the product of the rates for the forward cycle starting from $C_1$ is 
\be
R_f = W_{1f}.W'_{1f}.W_{2f}.W'_{2f}.W_{3f}.W_{4f} = \frac{1}{216 N_k^6}, 
\ee
while the product of the rates for the reverse cycle is 
\be
R_b = W_{1b}.W'_{1b}.W_{2b}.W'_{2b}.W_{3b}.W_{4b} =\frac{1}{192 N_k^6}.
\ee
Since $R_f \ne R_b$, it violates  the Kolmogorov criterion for the dynamics to obey detailed  balance~\cite{kolmogoroff1936theorie,1979-kelly,ederer2007thermodynamically}. 
Thus the steady state obtained using this diffusion kinetics \cite{2017-ltgv-pre-diffusiondriven} is  a nonequilibrium steady state. This argument, which we demonstrated explicitly for $k=2$ can be easily generalized for any larger $k$ with the introduction of $k-1$ primed configurations between 
$C_1 \rightleftharpoons C_2$ and another $k-1$ configurations between $C_2 \rightleftharpoons C_3$. 
The primed configurations differ by a lateral shift of a $k$-mer by one lattice unit in each step. 

\subsection{EDK model: Equilibrium steady state}

\begin{figure}
\centering
\includegraphics[width=\columnwidth]{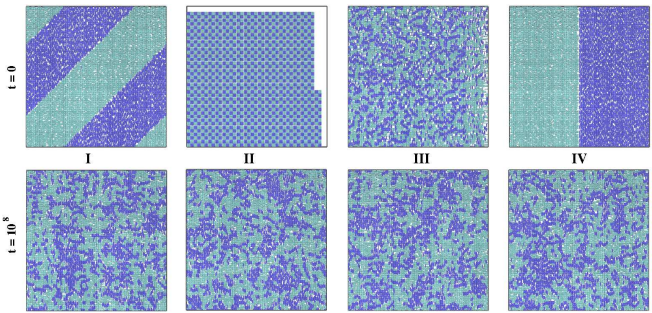}
\caption{Equilibrium steady states using EDK dynamics for $k=6$ for four different initial conditions (shown in top panels) -- (I) diagonally phase segregated, 
(II) checkerboard configuration, (III) spatially random, and (IV) vertically phase segregated.  The configuration
after $10^8$ Monte Carlo steps is shown in the bottom panels. Horizontal rods are coloured in cyan (light gray) while vertical rods are coloured in 
blue (dark gray). Empty sites are shown as white.}
\label{Fig2}
\end{figure}

We study numerically the model with EDK dynamics. This dynamics obeys detailed balance and we expect the system to equilibrate unless
ergodicity is broken. We first discuss the equilibrium steady state obtained using the EDK model for $k=6$. In Fig.~\ref{Fig2}, 
in the top panel we show four different initial configurations of the system -- (I) diagonally phase segregated, (II) checkerboard with the smallest block of any one orientation being of size $k \times k$, (III) spatially random, and (IV) vertically phase segregated. The bottom panels show the configurations achieved after $t=10^8$ Monte Carlo steps. The steady states obtained for the different initial conditions are statistically indistinguishable, and do not show any macroscopic
phase separation.  In fact, configurations (I) and (IV) are most striking, where despite starting from a macroscopically ordered state, the system 
randomizes in the equilibrium steady state. This result is consistent with earlier studies in the grand canonical ensemble of this system which found that the isotropic-nematic transition does not exist for $k < 7$~\cite{2007-gd-epl-on,2014-kr-pre-phase,2015-kr-epjb-phase}. Our simulation results  are in contradiction to the the phase segregated steady state 
obtained for the NEDK-R model for  $k=6$~\cite{2017-ltgv-pre-diffusiondriven}. We conclude that the results obtained in Ref.~\cite{2017-ltgv-pre-diffusiondriven} are due to the nonequilibrium nature of the dynamics and do not have any relevance for the corresponding equilibrium problem.

\begin{figure}
\centering
\includegraphics[width=\columnwidth]{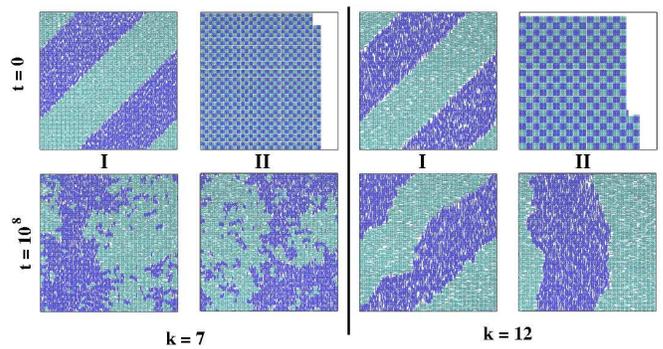}
\caption{Equilibrium steady states using EDK dynamics for $k=7$ and $k=12$ for two different initial conditions (shown in top panels) -- 
(I) diagonally phase segregated, and
(II) checkerboard configuration.  The configuration
after $10^8$ Monte Carlo steps is shown in the bottom panels. Horizontal rods are coloured in cyan (light gray) while vertical rods are coloured in 
blue (dark gray). Empty sites are shown as white.}
\label{Fig3}
\end{figure}

This naturally leads us to the question of whether a nematically phase segregated state exists in the micro canonical ensemble with EDK dynamics for 
$k \ge 7$, as would be expected from the studies in the grand canonical ensemble in equilibrium. In Fig.~\ref{Fig3}, the steady states obtained from two different initial configurations are shown for $k = 7$ and $k = 12$. For both $k=7$ and $k=12$, there exists a macroscopically phase segregated state. 
The phase boundaries between the horizontal and vertical $k$-mers are much sharper for $k=12$ in comparison to $k=7$. We also note that there is no preferred orientation of the phase boundaries. This is again in contrast to the results reported in Ref.~\cite{2017-ltgv-pre-diffusiondriven}, which observed a diagonally striped state in the long time limit, thus showing that the preferred orientation of domain walls is again a consequence of the nonequilibrium nature
of the dynamics.

\subsection{Nonequilibrium steady state}

We now ask whether the results diagonal striped phases for $k \ge 6$ obtained in Ref.~\cite{2017-ltgv-pre-diffusiondriven} using the NEDK-R algorithm
is true for generic nonequilibrium dynamics. Detailed balance condition may be violated in diverse ways.
We now study whether the nematically phase segregated state persists for $k=6$ when the system is evolved using the NEDK-CAc dynamics. 
If an ordered steady state exists, we wish to investigate whether the interface between the segregated states is along the diagonal.

\begin{figure}
\centering
\includegraphics[width=0.8\columnwidth]{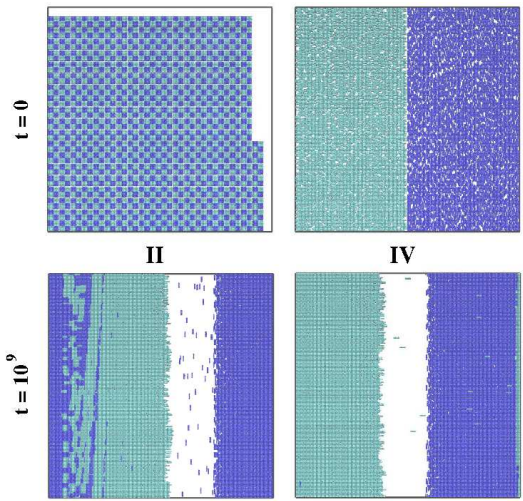}
\caption{Nonequilibrium steady states using NEDK-CAc dynamics for $k=6$  for two different initial conditions (shown in top panels) --
(II) checkerboard configuration,  and (IV) vertically phase segregated.  The configuration
after $10^9$ Monte Carlo steps is shown in the bottom panels. Horizontal rods are coloured in cyan (light gray) while vertical rods are coloured in 
blue (dark gray). Empty sites are shown as white. }
\label{Fig4}
\end{figure}

In Fig.~\ref{Fig4}, results for the steady state for $k=6$ using the NEDK-CAc dynamics are shown for two different initial conditions. The steady
states were obtained by running the simulations for $10^9$ Monte Carlo steps.
As can be seen from the snapshots, there exists a phase segregated state. Interestingly, it has two notable differences compared to the steady state obtained using the NEDK-R algorithm. The nematic phase segregated state now has vertical domain walls, with the the vertical rods to the left of the domain wall. Moreover, the stripes are compact, such that the voids also phase separate to form a strip (shown in white) different from the vertical-rich and 
horizontal-rich phases. 

\section{Discussion \label{sec:conclusion}}

To summarize, we studied the steady states obtained for the model of hard rods on a square lattice using three different kinetic rules. Each
of these rules consist of local translation of $k$-mers keeping the total density as well as the nematic order parameter fixed. The primary
motivation of the paper was to re-examine the results obtained for the NEDK-R algorithm in Ref.~\cite{2017-ltgv-pre-diffusiondriven},
where systems with $k\geq 6$ were shown to phase separate with the domain walls separating the horizontal-rich and vertical-rich
phases being oriented diagonally. Similar results were also reported in another subsequent publication \cite{2017-tlbl-arxiv-pattern}. These results were in contradiction to grand canonical simulations of the same system where
there is nematic ordering only for $k\geq 7$, and questioned the well-established equivalence of ensembles for determining 
the macroscopic features of systems in equilibrium. We resolved this ambiguity by showing that the diffusion dynamics 
studied in Ref.~\cite{2017-ltgv-pre-diffusiondriven} violates detailed balance condition and drives the system into a nonequilibrium
steady state. Therefore, the results obtained using the NEDK-R algorithm have no relevance for the equilibrium phases. 

In addition, we showed that when the diffusion move satisfies detailed balance (EDK dynamics), the system does not get trapped in
any nonequilibrium metastable state, and relaxes to its true equilibrium configuration: disordered for $k=6$ and phase segregated
for $k \ge 7$ (as demonstrated for $k=7$ and $k=12$). Thus the current results are consistent with those obtained in earlier simulations 
in the grand canonical ensemble, confirming the expected equivalence of micro canonical and grand canonical ensembles in equilibrium.
Also, the domain walls in the phase segregated regime (for $k \ge 7$) could orient in directions different from the diagonal direction showing
that the diagonal domain walls obtained using NEDK-R algorithm in Ref.~\cite{2017-ltgv-pre-diffusiondriven} are not generic.

There is no canonical choice of dynamics that violate detailed balance. To show that very different steady states can be obtained
by altering the dynamics, we studied the steady states of the system using NEDK-CAc dynamics. 
Using this dynamics we obtained phase segregation for $k=6$ where in addition to the horizontal and vertical $k$-mers, the voids also
phase separate.  Also, the domain walls are vertical rather than horizontal, showing that there may be interesting specific patterns that arise
when the diffusive dynamics is altered. If the choice of the nonequilibrium dynamics is motivated by some physical system, then the corresponding 
ordered pattern may acquire significance in the context of that specific system.

\section*{Acknowledgement}

MKM acknowledges financial support from Ramanujan Fellowship, DST, INDIA (13DST052) and the IIT Bombay Seed Grant (14IRCCSG009).

%
\end{document}